\documentclass[twocolumn]{aastex62}

\newcommand{\rsub}{r_\mathrm{sub}}
\newcommand{\pdv}{\partial}

\newcommand{\ledd}{\ell_\mathrm{Edd}}

\received{August 2, 2019}
\revised{September 12, 2019}
\accepted{September 16, 2019}
\shorttitle{Redefining the torus}
\shortauthors{S. F. H\"onig}
\begin{document}

\title{Redefining the torus: A unifying view of AGN in the infrared and sub-mm}

\correspondingauthor{Sebastian F. H\"onig}
\email{s.hoenig@soton.ac.uk}

\author[0000-0002-6353-1111]{Sebastian F. H\"onig}
\affil{School of Physics \& Astronomy, University of Southampton, Southampton SO17 1BJ, United Kingdom}



\begin{abstract}
The advent of high-angular resolution IR and sub-mm interferometry allows for spatially-resolved observations of the parsec-scale environment of active galactic nuclei (AGN), commonly referred to as the ``torus.'' While molecular lines show the presence of large, massive disks, the IR observations appear to be dominated by a strong polar component that has been interpreted as a dusty wind. This paper aims at using characteristics shared by AGN in each of the wavebands and a set of simple physical principles to form a unifying view of these seemingly contradictory observations: Dusty molecular gas flows in from galactic scales of $\sim$100\,pc to the sub-parsec environment via a disk with small to moderate scale height. The hot, inner part of the disk puffs up due to IR radiation pressure and unbinds a large amount of the inflowing gas from the black hole's gravitational potential, providing the conditions to launch a wind driven by the radiation pressure from the AGN. The dusty wind feeds back mass into the galaxy at a rate of the order of $\sim0.1-100\,M_\sun$/yr, depending on AGN luminosity and Eddington ratio. Angle-dependent obscuration as required by AGN unification is provided by a combination of disk, wind, and wind launching region.
\end{abstract}



\section{Introduction} \label{sec:intro}

The active growth of supermassive black holes occurs via accretion of dust and gas in the nuclei of galaxies. The standard model of unification of active galactic nuclei (AGN) posits that the parsec-scale dusty molecular structure of the accretion process forms a geometrically thick entity -- dubbed the ``torus'' -- around the central X-ray/UV/optical emission source, providing the angle-dependent obscuration to explain the difference between type 1 (unobscured) and type 2 (obscured) AGN \citep[e.g.][]{Ant85,Ant93,Urr95,Ram17}. While the ``torus'' was originally introduced for its obscuring nature to optical emission \citep{Ant85}, it has since become a catch phrase for dense, dusty molecular gas present in the tens of parsec-scale environment of the AGN \citep[e.g.][and references therein]{Ram17,Com19}.

In recent years, high-angular resolution observations of local Seyfert galaxies in the infrared (IR) with the Very Large Telescope Interferometer (VLTI) showed that the dust on these scales is not distributed in one single, toroidal structure \citep{Hon12,Hon13,Tri14,Lop14,Lop16,Lef18}. Instead, simultaneous radiative transfer modelling of IR interferometry and the IR spectral energy distribution (SED) implies a two-component structure with an equatorial, thin disk and a polar-extended feature, which may originate from a dusty wind forming a hollow cone and defining the edges of the narrow-line region \citep{Hon17,Sta17}. This is contrary to classical thick torus models that assume that the obscuring and emitting mass is distributed either smoothly or in clumps in a single toroidal structure extending from sub-parsec to tens of parsec scales.

In parallel, the Atacama Large sub-Millimeter Array (ALMA) probed the molecular phase of the dusty gas on similar spatial scales in several local AGN \citep[e.g.][]{Gal16,Ima16,Gar16,Alo18,Alo19,Com19}. These observations found nuclear rotational structures in several molecular lines (e.g. CO, HCN, HCO$^+$) in most of these galaxies, with potential outflow signatures present in some of them. Those ``molecular tori'' are about a factor of 10 larger than the IR emission sizes (several 10\,pc in the sub-mm) and almost exclusively located in the plane of the accretion disk. Taken at face value, those observations are qualitatively more consistent with a single ``geometric torus''.

Fundamentally, both IR and sub-mm observations trace the same gas flow. Hence, both sets of data, while sometimes considered contradictory, should be governed by the same physics. This paper takes a look at the structures that can be inferred empirically from the observations in the IR and sub-mm and applies some fundamental physical principles to unify the observations at both wavelength regimes. This work will focus on observations of radio-quiet AGN in the Seyfert-luminosity regime, but relations to higher luminosity sources will be discussed where applicable.

\section{The infrared continuum view: Dusty disk and polar elongation}\label{sec:ir_cont_view}

\subsection{Basic emission characteristics}\label{subsec:basics}

The near- and mid-IR emission from AGN usually presents itself as a point source in single-dish 8m-class telescope data \citep[for a comprehensive summary see][]{Asm14}, with notable exceptions that will be discussed below \citep[e.g.][]{Boc00,Pac05,Reu10,Hon10a,Asm16,Asm19}. For nearby Seyfert galaxies ($D_A < 100\,$Mpc), this corresponds to scales of $<$30\,pc at 2.2\,$\micron$ and $<$180\,pc at 12\,$\micron$. Several authors have compiled SEDs for Seyfert AGN at those resolutions, which will be summarised in the following \citep[e.g.][]{Ram09,Hon10a,Alo11,Esq14}.

In general, the IR SED of Seyferts is characterised by a well-documented rise from about 1\,$\micron$ towards longer wavelengths. The sharp and universal turn at $\sim$1\,$\micron$ has its origin in the fact that dust will sublimate when heated above a temperature of $1200-1900\,$K, depending on dust species \citep[e.g.][]{Bar87,Net15}. Dust sublimation introduces a characteristic sublimation radius $\rsub$ marking the inner boundary of the dust distribution (see Sect.~\ref{subsec:ir_intf} for a more detailed description). As predicted by the local thermal equilibrium equation and as observationally confirmed \citep[e.g.][]{Sug06,Kis11a,Kos14}, $\rsub$ scales with the square-root of the AGN luminosity, $\rsub \propto L^{1/2}$.

The difference between obscured type 2 sources and unobscured type 1 AGN is most prominent in the $3-5\,\micron$ range, with the hot dust emission significantly suppressed in type 2s due to self-absorption of the obscuring structure. In addition, several authors point out an excess of hot dust emission in unobscured AGN above what single-structure radiative transfer models predict for full IR SED fits \citep[e.g][]{Mor09,Alo11}. This $3-5\,\micron$ excess, or ``bump'' as it may display a local or global maximum in the SED in some sources, has been interpreted as a separate emission component with its strength varying from source to source. In contrast, the mid-IR luminosity is of similar magnitude in both obscured and unobscured AGN when compared to the intrinsic AGN luminosity, with the anisotropy as low as a factor of $\sim$1.5 at 12\,$\micron$. \citep[e.g.,][]{Hor08,Gan09,Ram11,Hon11,Asm11,Asm15}. This still holds when accounting for anisotropy of the primary accretion disk radiation \citep[e.g.][]{Sta16}.

For dusty, molecular gas accreted from the host galaxy, it is expected that the silicates produce an absorption or emission feature at $\sim$10\,$\micron$, depending on the line-of-sight to the hottest dust. While some obscured AGN show strong silicate absorption features as expected, the corresponding silicate emission features in unobscured AGN from the hot dust are very shallow \citep{Hao07,Hon10a,Alo11,Gar17}. Indeed, the strongest absorption features may be due to dust in the host galaxy and not related to the AGN environment \citep{Gou12}, meaning that the silicate absorption originating from the nuclear environment is rather weak.

\subsection{IR interferometry}\label{subsec:ir_intf}

About 40 nearby AGN have been observed with IR interferometry in the near-IR or mid-IR. In the near-IR, observations are in general agreement with the hot dust emission emerging from close to the dust sublimation region, showing the expected $\rsub \propto L_\mathrm{AGN}$ scaling \citep[e.g.][]{Swa03,Kis07,Kis09b,Pot10,Kis11a,Wei12}. However, both near-IR interferometry and reverberation mapping \citep[e.g.][]{Sug06,Kos14} find that the absolute scaling of the size-luminosity relation is offset towards smaller sizes by a factor of 2$-$3 with respect to predictions based on mixed-dust opacities in line with standard ISM dust for an average size of 0.07\,$\micron$. The observations can be reconciled with theory when assuming that the hot dust is primarily comprised of large graphite grains \citep[e.g.][]{Kis07,Hon17}. This is supported by the observed near-unity dust surface emissivities determined from near-IR interferometry \citep{Kis11a,Kis11b} and follows naturally from differential dust sublimation and the fact that the 2.2\,$\micron$ emission traces temperatures that can only be survived by the largest grains \citep{Gar17,Hon17}.

In the mid-IR, the picture is more complex: Most sources show a drop in visibility from unity to between 0.1$-$0.8 on shorter baselines of 40--60\,m. On longer baselines up to 130\,m, the visibility remains essentially unchanged with respect to the short baselines, indicating that some of the emission stays unresolved \citep[e.g.][]{Bur13,Lop16,Lef18,Lef19}. Without further information, the emission is commonly interpreted by means of a two-component model, consisting of a resolved and unresolved emission source.

In a few sources, it has been found that the resolved mid-IR component is elongated in the rough direction of the polar direction of the AGN \citep{Hon12,Hon13,Tri14,Lop14}. More precisely, the mid-IR emission is extended towards the edge of the ionisation cones \citep[e.g][]{Lop16,Sta17,Lef18,Sta19}. The extended emission accounts for a significant part of the point-like source seen in most single-telescope mid-IR imaging data, reaching 60-80\% in the interferometrically best-covered AGN in NGC1068, the Circinus galaxy, and NGC3783. \citet{Asm16} report that in 21 nearby AGN such ``polar'' features can be traced out to 100s of parsecs, and the detection frequency is presumably limited not by occurance of these features but by surface brightness sensitivity \citep[see][]{Asm19}. They speculate that these very extended features are the continuation of the 0.1$-$1\,pc polar emission structures seen by IR interferometry.

Circinus and NGC1068 are close and bright enough that the mid-IR emission source can be resolved below visibilities of 0.1 \citep{Tri14,Lop14}. At these levels, the otherwise unresolved component is partially resolved as a geometrically thin disk approximately parallel to the maser disk seen in both objects (with a potential third emission source at low flux contribution present in both sources). The maximum size of the disks are smaller than the size of the polar-extended emission (factor 1.6 in Circinus, 9 in NGC1068) and their axis ratios indicate a high inclination (major:minor axis ratio of $>$2:1 in Circinus and 6:1 in NGC1068). By extrapolation, the interferometrically unresolved mid-IR emission seen in more distant AGN may well be the disk component that is partially resolved in NGC1068 and Circinus.

It is worth noting that silicate emission or absorption features are mostly absent from the interferometric data, or, at least, not more pronounced than in unresolved, single-telescope data. The notable exception to this is NGC1068, where the spectrally-resolved visibilities show a deep silicate absorption feature at about 9.7$\,\micron$. Such a behaviour would be expected for a clumpy torus seen under type-2-like inclination, where self-absorption due to higher opacities in the feature as compared to the continuum causes the source to appear larger \citep{Hon10b}. That no other AGN shows a similar behaviour, contradicts this explanation as a genuine feature in AGN.

\subsection{Modelling the resolved and unresolved IR emission}\label{subsec:model_ir}

The size and magnitude of the polar emission components in both type 1 and type 2 AGN are difficult to reconcile with torus models using a single toroidal mass distribution, unless very specific geometries and line-of-sights across all sources are assumed. Hence, new radiative transfer models consisting of a disk and a polar component have been developed \citep[e.g][]{Gal15,Hon17,Sta17}. In line with observations, these models assume that the polar component originates from a hollow cone at the edges of the ionisation region. They successfully and simultaneously reproduced the total and resolved emission in NGC3783 \citep{Hon17} and the Circinus galaxy \citep{Sta17,Sta19}. In addition, \citet{Hon17} noted that such models are able to qualitatively and quantitatively reproduce the distinct $3-5\,\micron$ bump in the SED of unobscured AGN, associating it primarily with the inner hot part of the disk component \citep[see also][]{Gon19}.

\subsection{The phenomenological picture in the IR}\label{subsec:pheno_ir}

\begin{figure}
\begin{center}
\includegraphics[width=\columnwidth]{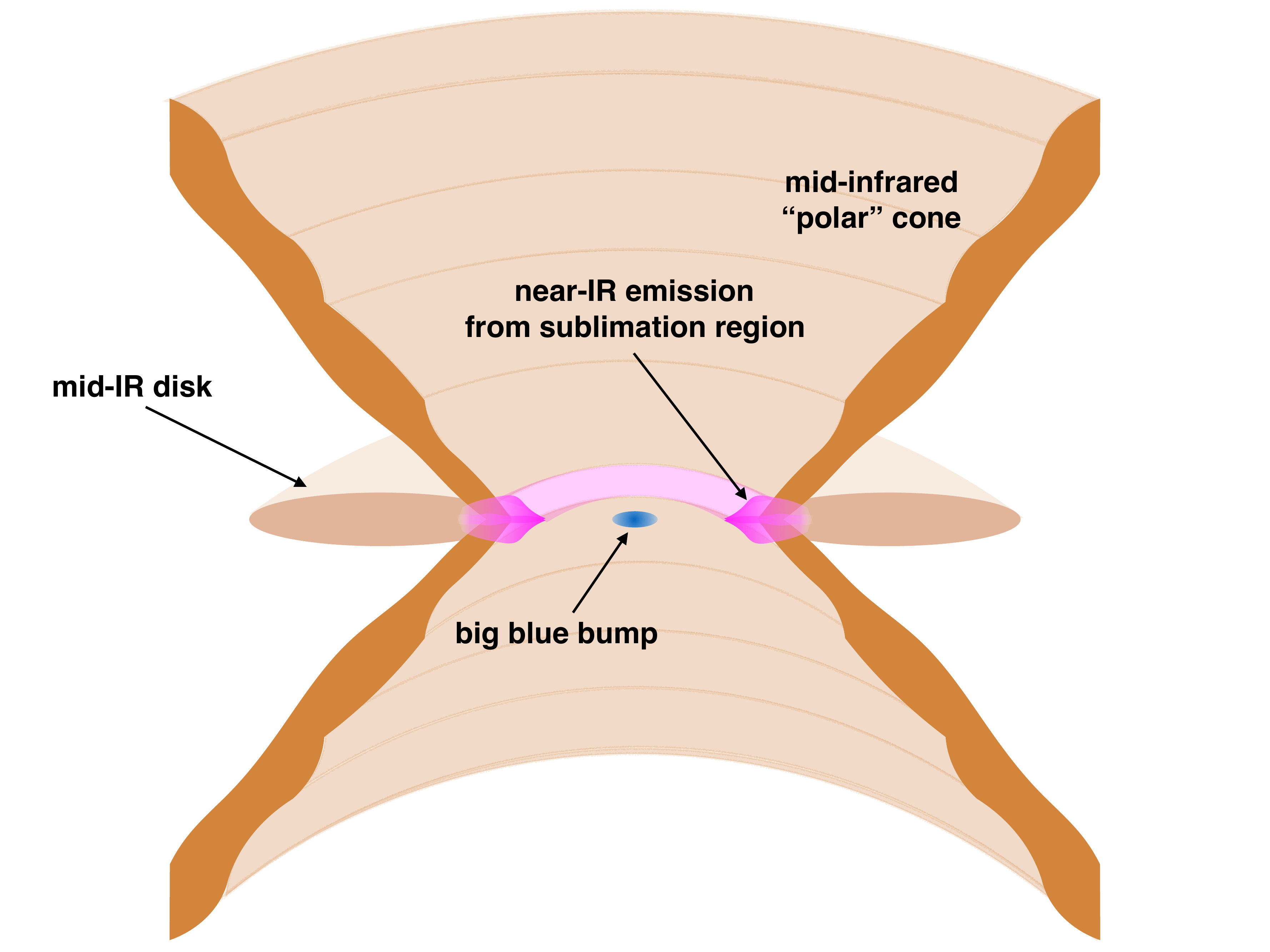}
\end{center}
\caption{\label{fig:uv_ir}
Schematic view of the pc-scale AGN infrared emission consisting of a geometrically-thin disk in the equatorial plane (light brown) and a hollow dusty cone towards the polar region (dark brown). The inner part of the disk (pink) emits the near-IR emission dominating the $3-5\,\micron$ bump.}
\end{figure}

Empirically, this leads to the conclusion that the infrared emission is originating from two components: a disk and a hollow cone (see Fig.~\ref{fig:uv_ir}). While the disk may dominate the near-IR emission in unobscured AGN, the bulk of the mid-IR emission emerges from a hollow dusty cone towards the edge of the high ionisation region. As pointed out in literature, in many AGN with confirmed polar mir-IR emission features, only one rim of the dusty cone seems to be detected \citep[e.g][]{Tri14,Lop14,Lop16}. \citet{Sta17} use detailed radiative transfer modelling to argue that such one-sided emission requires both anisotropic emission of the big blue bump as expected for an accretion disk \citep{Net87} as well as a misalignment between the axis of the accretion disk and the axis of the hollow cone. Alternatively, partial obscuration of one rim by the host galaxy due to a misalignment of AGN and galaxy rotation axis may contribute. A physical interpretation of the two-component structure will be discussed in Sect.~\ref{sec:phys_desc}.

\section{The molecular view: disks}\label{sec:mol_view}

\subsection{CO lines observed with ALMA}\label{subsec:co_alma}

While sub-mm observations of AGN have been previously performed \citep[e.g.][]{San89,Bar89,Pla91,Jac93,Tac94,Tac99,Sch99,Sch00b,Dav04,Kri05,Cas08}, ALMA revolutionised spatial resolution, $uv$-plane, and frequency coverage. Several authors have presented results for nearby ($D<50$\,Mpc) Seyfert and low-luminosity AGN with spatial resolution of the order of a few to 10\,pc, i.e. only slightly lower than the VLT interferometer in the IR \citep[e.g.][]{Com13,Com14,Gar14,Ima16,Gal16,Gar16,Aud17,Ima18,Alo18,Com19,Alo19}. Most of these observations specifically target various rotational bands of CO, HCN, and HCO$^+$, tracing gas densities of a few $\times 10^4$\,cm$^{-3}$ to $10^6$\,cm$^{-3}$.

\begin{table*}
\begin{center}
\caption{\label{tab:scaleheight} \textup{
Kinematic properties and derived scale heights for sub-mm CO and infrared H$_2$ emission lines.
}
}
\begin{tabular}{lccccl}
\hline\hline
Object & line & $v_\mathrm{rot}\,^a$ & $\sigma\,^b$ & $H/R\,^c$ & reference \\
& & (km/s) & (km/s) & & \\ \hline
\multicolumn{6}{c}{sub-mm CO lines} \\ \hline
NGC3227\,$^d$ & CO(2--1), CO(3--2) & 150--200 & 20--30 & 0.1--0.2 & \citet{Alo19} \\
NGC5643 & CO(2--1) & 115 & 60 & 0.52 & \citet{Alo18} \\
NGC1365 & CO(3--2) & 95--187 & 35 & 0.19--0.37 & \citet{Com19} \\
NGC1566 & CO(3--2) & 134--480 & 40 & 0.08--0.30 & \citet{Com19} \\
IC5063  & CO(2--1), CO(4--3) & 175--250 & 50 & 0.20--0.28 & \citet{Das16} \\
Circinus & CO(3--2) & 70 & 35 & 0.5 & \citet{Kaw19} \\
NGC1068 & CO(6--5) & 70 & 10 & 0.14 & \citet{Gal16} \\ \hline
\multicolumn{6}{c}{infrared H$_2$ lines} \\ \hline
NGC3227 & H$_2$(1--0)S(1) & 65 & 90 & 1.38 & \citet{Hic09} \\
NGC3783 & H$_2$(1--0)S(1) & 26 & 33 & 1.27 & \citet{Hic09} \\
NGC4051 & H$_2$(1--0)S(1) & 37 & 44 & 1.19 & \citet{Hic09} \\
NGC4151 & H$_2$(1--0)S(1) & 82 & 67 & 0.82 & \citet{Hic09} \\
NGC6814 & H$_2$(1--0)S(1) & 23 & 43 & 1.87 & \citet{Hic09} \\
NGC7469 & H$_2$(1--0)S(1) & 38 & 63 & 1.66 & \citet{Hic09} \\
Circinus & H$_2$(1--0)S(1) & 36 & 51 & 1.42 & \citet{Hic09} \\
NGC1068 & H$_2$(1--0)S(1) & 21 & 102 & 4.85 & \citet{Hic09} \\
\hline
\end{tabular}
\end{center}
\textit{--- Notes:}$^a$\,deprojected rotational velocity  except for the CO data of IC5063, Circinus, and NGC1068 where no inclination was given;  $^b$\,velocity dispersion of the molecular gas; $^c$\, scale height estimate for the rotational disks , $H/R \approx \sigma/v_\mathrm{rot}$; $^d$\,model-inferred values after accounting for outflow and bar motion, as given in Sect.~5.3 in \citet{Alo19}.
\end{table*}

The kinematics observed in the molecular emission lines can be complex, with influences from the host galaxy on larger scales ($\ga$100\,pc, e.g. via bars or other dynamic resonances) and rotation and outflows dominating smaller scales \citep[e.g.][]{Ima16,Gal16,Gar16,Aal16,Aud17,Aal17,Alo18,Com19,Alo19}. The central 30$-$50\,pc in Seyfert nuclei recently observed with ALMA do have in common that they show a clear rotational pattern for the bulk of the CO gas\footnote{At the time this was written, the various rotational levels of CO had the best observational coverage across local Seyfert galaxies. When comparing CO to other molecular lines, e.g HCN, in some of the publications, the overall kinematic properties are similar, though the spatial distribution may vary (see Sects.~\ref{subsubsec:thinmoldisk_rad} \& \ref{subsec:uni_colddisk}). Hence CO is used as a proxy here.}. An overview of extracted nuclear disk properties of Seyfert-type AGN are listed in Table~\ref{tab:scaleheight}. The rotational velocities, $v_\mathrm{rot}$, and gas velocity dispersion, $\sigma$, have either been directly given in the referenced papers or extracted from the moment maps and pv-diagrams. Excluded are objects where the ALMA data has resolution $\ga50$\,pc, where the publication does not convey sufficient kinematic information to determine the scale height, that do not unequivocally show AGN activity in optical \citep{Ver10} and X-ray observations \citep{Bau13}, or that have an implied level of activity that would not typically identify them as ``normal'' Seyfert AGN.

Assuming a disk in hydrostatic equilibrium (see Sect.~\ref{subsubsec:thinmoldisk_vert}), it is possible to estimate the scale height as $H/R \approx \sigma/v_\mathrm{rot}$. The corresponding values for the CO emission in local Seyfert galaxies are also shown in Table~\ref{tab:scaleheight}. Interestingly, none of these molecular disks are geometrically thick. The typical scale heights of the CO gas in those ALMA observations is $H/R\sim0.15-0.3$. 

The required scale height of a single obscurer in terms of the torus picture can be inferred from number ratios of X-ray obscured (=type 2) and unobscured (=type 1) sources assuming that the X-ray obscuring gas is located within the ``torus'' region. Using the X-ray background analysis from \citet{Ued14}, the ratio between type 1 ($N_H < 10^{22}$\,cm$^{-2}$) and type 2 ($N_H \ge 10^{22}$\,cm$^{-2}$) sources can be inferred in the range of 1:1-2.3 for AGN with Seyfert luminosities of $\sim10^{42}-10^{45}$\,erg/s. This corresponds to an X-ray covering factor of $\sim0.3-0.7$\footnote{\citet{Ric17} showed that the X-ray covering factor in hard X-ray-selected AGN is Eddington-ratio dependent. For Eddington ratios consistent with Seyfert-type AGN in the range of $\ell_\mathrm{Edd}\sim0.01-0.2$\, the observed covering factor is $\sim$0.3$-$0.8, consistent with the \citet{Ued14} result once the Eddington ratio distribution is factored in.}. From simple geometric considerations, $C = (H/R)/\sqrt{1+(H/R)^2)}$, which implies that the observed CO gas has a covering factor of only $C_\mathrm{CO} \sim 0.2-0.3$. 

\subsection{H$_2$ in the near-IR and H$_2$O masers}\label{subsec:h2_H2o}

Similar spatial scales as with ALMA are reached in the near-IR for the H$_2$ molecule. \citet{Hic09} report kinematic modelling of the near-IR H$_2$(1--0)S(1) ro-vibrational transition at $2.212\,\micron$, which will be referred to simply as H$_2$ through the rest of the paper. This emission line traces hotter gas than the sub-mm CO rotational lines --- 1000--2000\,K as compared to 20--50\,K. Deprojected rotational velocities and velocity dispersions at 30\,pc from the nucleus from \citet{Hic09} and the derived scale heights for a sample of local Seyfert galaxies are listed in Table~\ref{tab:scaleheight}. The H$_2$ emission seems vertically more extended than the CO lines due to its higher observed velocity dispersion, with typical $H/R \sim 1.2-1.4$. \citet{Hic09} note that due to the observed co-planar rotation, warps cannot be responsible for the observed high $\sigma$ values. The authors further point out that the velocity dispersion of H$_2$ is significantly larger than the stellar velocity dispersion in the same region, implying that the stellar cluster cannot account for the observed turbulence. As a result, \citet{Hic09} conclude that the H$_2$ must form a geometrically thick disk, potentially inflated by starformation near the nucleus. Converting $H/R$ into an H$_2$ covering factor leads to $C_{\mathrm{H}_2} \sim 0.77-0.81$, i.e. almost a spherical distribution. 

It is important to note that separating outflows from disk components is not straight-forward. Due to conservation of angular momentum, outflows can have a rotational component that can be mistaken for disk rotation. Hence, part of the comparably large $\sigma(\mathrm{H}_2)$ may be due to a combination of a disk and an outflow. The kinematically estimated $H/R$ for H$_2$ and its corresponding $C_{\mathrm{H}_2}$ should, therefore, be considered an upper limit. Nevertheless, it is likely that the H$_2$ disks have a larger $H/R$ than disks seen in sub-mm molecules.

Finally, NGC 1068 and Circinus are well-known for their nuclear maser disks \citep[e.g.][]{Gre97,McC09}. These features are usually seen on parsec scales, i.e. at about the same scale as the VLTI observations and slightly smaller than ALMA. H$_2$O masers trace the densest gas with $n_\mathrm{H}\ga10^9$\,cm$^{-2}$. After accounting for warps, the geometric thickness of those maser disks is very small.

\subsection{The phenomenological picture in the sub-mm}\label{subsec:pheno_submm}

\begin{figure}
\begin{center}
\includegraphics[width=\columnwidth]{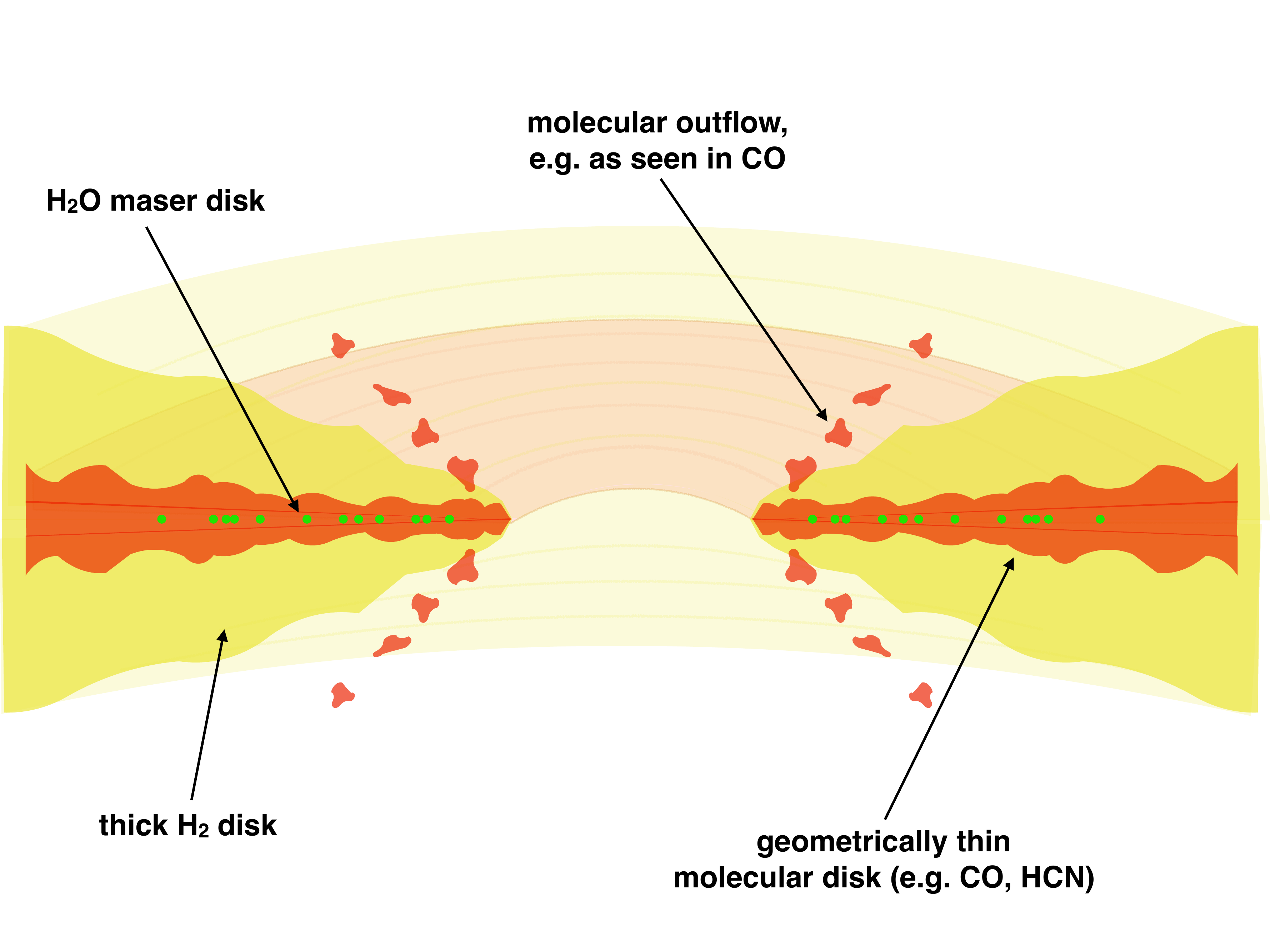}
\end{center}
\caption{\label{fig:uv_submm}
Schematic view of the AGN environment from molecular emission lines. The high-density masers (green) are located in the equatorial plane with CO/HCN (red) forming a thin disk and H$_2$ (yellow) a thick disk.}
\end{figure}

Fig.~\ref{fig:uv_submm} puts together the empirical view of the circumnuclear environment of radio-quiet Seyfert AGN from the perspective of the observed molecular emission lines. The vertical distribution of gas shows a clear density and temperature stratification. Qualitatively, such a profile is expected for a disk in hydrostatic equilibrium where the densest material sinks to the plane of the disk. The schematic view also accounts for the nuclear molecular outflows seen in at least some AGN \citep[e.g.][]{Gal16,Alo19}. As discussed in Sect.~\ref{subsec:mdotw}, some AGN show more jet-like outflow features. These are usually seen in more radio-loud/intermediate objects and may be brought into the presented schematic by adding collimated molecular emission close to rotation axis.

\section{A physical description of the mass distribution}\label{sec:phys_desc}

\subsection{Physical conditions of the dusty molecular gas}\label{subsec:phys_cond}

The previous sections demonstrated that the observed emission in the IR dust continuum and molecular lines show very different structures. To form a unifying view of the mass distribution, it is necessary to take into account some general physical principles, which are well established in literature based on theory or observations.

\paragraph{1. The IR dust continuum delineates the $\tau_\mathrm{IR} \sim 1$ surface of the dust distribution and may not be representative of the entire mass distribution.} This is a result of the radiative transfer equation as, in a simplified version, the emitted intensity $S_\lambda$ can be approximated by $S_\lambda \propto \tau_\lambda B_\lambda$, where $\tau_\lambda$ is the optical depth at wavelength $\lambda$ and $B_\lambda$ is the Planck function. Therefore, the largest emission contribution will come from $\tau_\lambda\sim1$\footnote{Note that this description is a simplification of the full radiative transfer and neglects self-absorption. For $\tau_\lambda>1$, the observed intensity will be $S_\lambda \approx B_\lambda\cdot\exp(-\tau_\lambda)$. Including this self-absorption effect, the emitting surface would be more correctly set at $\tau_\lambda = 2/3$.}. The observed ``surface'' of a mass distribution is, therefore, wavelength dependent, which means that different wavelengths will trace different regions of the mass distribution.

\paragraph{2. For typical conditions in the circumnuclear environment, the gas and dust will not have the same temperature, i.e. the gas temperature $T_\mathrm{g}$ is greater than the dust temperature $T_\mathrm{d}$.} \citet{Wil19} show from their radiation-hydrodynamical (RHD) simulations that even if dust and gas are hydrodynamically coupled, the temperature may be quite different. From their Fig.~9, an approximate relation of $$\log T_\mathrm{g} \approx 1.7\cdot \log T_\mathrm{d} - 0.25$$ can be derived for non-shocked dusty gas. This means that hot gas with $T_\mathrm{g} \sim 4500$\,K can be co-spatial with dust radiating in the mid-IR at $T_\mathrm{d} \sim 200$\,K. 

\paragraph{3. As the dust opacity $\kappa_\mathrm{d} \sim 10^{2-3}\cdot\kappa_\mathrm{g}$ is much greater than the gas opacity\footnote{Here, the term ``gas opacity'' refers to the opacity of ionised Hydrogen gas as used in the Eddington limit, $\kappa_\mathrm{g}=\sigma_\mathrm{Thom}/m_p$, defined by the Thomson cross-section $\sigma_\mathrm{Thom}$ to electron scattering and the proton mass $m_p$.} $\kappa_\mathrm{g}$, AGN with Eddington ratios $\ledd \ga 0.01$ will inevitably drive dusty winds by radiation pressure.} Several authors have noted this fact for the direct UV/optical AGN emission \citep[e.g.][]{Pie92,Hon07,Fab09,Ric17} and RHD simulations confirm it for both single dusty clumps as well as massive dusty molecular disks \citep[e.g.][]{Dor11,Sch11,Wad12,Nam14,Nam16,Cha17,Wil19}. Such radiation pressure driving is most effective for regions with $\tau_V\sim1$. It is important to note that this principle also applies to other wavelengths where the AGN emits a significant fraction of its overall luminosity. As such, dusty gas with near-IR optical depth of at least $\tau_\mathrm{NIR}\sim1$ will also contribute to wind driving \citep[e.g.][]{Kro07,Cha17,Ven19}. As a consequence, IR radiation pressure will affect the distribution of dusty gas \citep{Kro07,Nam16}.

\paragraph{4. Absent significant vertical pressure, a geometrically thick configuration $H/R\sim1$ cannot be easily maintained.} While being a trivial statement from a theoretical point of few, the practical consequence of it is that the simplest approaches to producing a geometrically thick torus have often failed. \citet{Kro88} already noted that for typical gas and dust temperatures observed in AGN, thermal pressure is too low to sustain the observed velocity dispersion. Alternatively, a clumpy medium requiring fully elastic cloud-cloud collisions \citep[e.g][see discussion in Sect.~\ref{subsubsec:thinmoldisk_vert}]{Kro88,Vol04,Bec04}, high supernova rates \citep[e.g.][]{Wad12}, turbulence from accretion shocks \citep[e.g.][]{Vol18} or warps \citep[e.g.][]{Sch00a} have been explored.

\subsection{A physical picture}\label{subsec:phys_pic}

The empirical picture laid out in Sects.~\ref{sec:ir_cont_view} \& \ref{sec:mol_view} implies the presence of a relatively thin disk and a polar hollow cone towards the edge of the ionisation region. This hollow cone may be interpreted as a dusty, molecular wind on parsec scales. The following will discuss physical mechanisms related to the disk and wind regions, based on the principles laid out in Sect.~\ref{subsec:phys_cond}. While the remainder of Sect.\ref{sec:phys_desc} will be based on theoretical arguments, Sect.~\ref{sec:unify} will discuss the implications of the emerging picture in the context of observations.

\subsubsection{The geometrically thin molecular disk: vertical density stratification of molecular lines}\label{subsubsec:thinmoldisk_vert}

The high-density molecular tracer lines imply that on pc scales (H$_2$O disk masers) to 10s pc scales (e.g. CO, HCN), the accretion flow is settled to a relatively thin disk. This molecular disk appears to be co-planar with the thin-disk-like mid-IR emission seen in Circinus and NGC1068. Following from principle 1, it is reasonable to assume that the observed disk-like mid-IR emission originates from the $\tau_\mathrm{MIR}\sim1$ surface of this dusty molecular disk \citep[see also][]{Sta19}. The mass in such a disk can be inferred to be $\ga$10\% of the mass of the black hole (see Sect.~\ref{subsec:uni_colddisk}), which make it at least vertically self-gravitating, although on 10s of pc scales, the total gravitational potential consists of a mix of black hole and nuclear star cluster potentials.

As per principle 4, the scale height of such cool disks is small. A commonly invoked physical model in these cases is an isothermal disk. Observations have shown that the cool, high-density tracer lines such as CO (few 10\,K) are about co-spatial with hotter, lower-density tracers such as H$_2$. Therefore, the dense disks are multi-phase media. Notwithstanding the choice of tracer, the vertical density distribution of an isothermal disks can be inferred from the hydrostatic equilibrium equation 
\begin{equation}\label{eq:hse}
\frac{\pdv P}{\pdv z} = -g_z \ \rho(z)
\end{equation}
with the pressure $P=\rho(z) \cdot k_B T_\mathrm{gas}/m_\mathrm{gas}$, gas temperature $T_\mathrm{gas}$, and gas particle mass $m_\mathrm{gas}$. Solving the differential equation provides the well-known exponential profile $\rho(z) \propto \exp(-z^2/2h^2)$ with the square of the scale height $(h/r)^2=k_B T_\mathrm{gas}/(m_\mathrm{gas} \cdot v_\mathrm{rot}^2)$. 
Such a vertical density structure implies that lower-density gas will appear thicker than higher-density gas. 

As already pointed out in Sect.~\ref{subsec:phys_cond}, for typical gas temperatures of H$_2$ or CO, the scale height of an isothermal disk would be vanishingly small, requiring either a hotter embedding medium ($T_\mathrm{gas}\sim10^6\,$K), or additional turbulence, e.g. from cloud-cloud collisions \citep[e.g.][]{Kro88,Vol04,Bec04}, which may be interpreted as a dynamical temperature. Following \citet{Vol04}, the scale height for a turbulent medium with clouds obeying the Jeans criterion (i.e. being stable, or, in the limit, marginally stable) can be calculated as
\begin{equation}
    h/r \le \frac{\sqrt{9kT_\mathrm{gas}}}{\Phi_V r m_\mathrm{gas} \sqrt{8Gn_H}},
\end{equation}
where $\Phi_V$ is the volume filling factor of the turbulent medium, and $n_H$ the Hydrogen number density of the gas in the cloud. In more convenient units, this resolves to $h/r \le 0.93 \cdot (T_{1000}/n_{H;5})^{1/2} \cdot r_{10}^{-1}$ with the gas temperature in units of 1000\,K, the Hydrogen number density in units of $10^{5}$\,cm$^{-2}$ and the distance from the centre in units of 10\,pc.

\begin{figure}
\begin{center}
\includegraphics[width=\columnwidth]{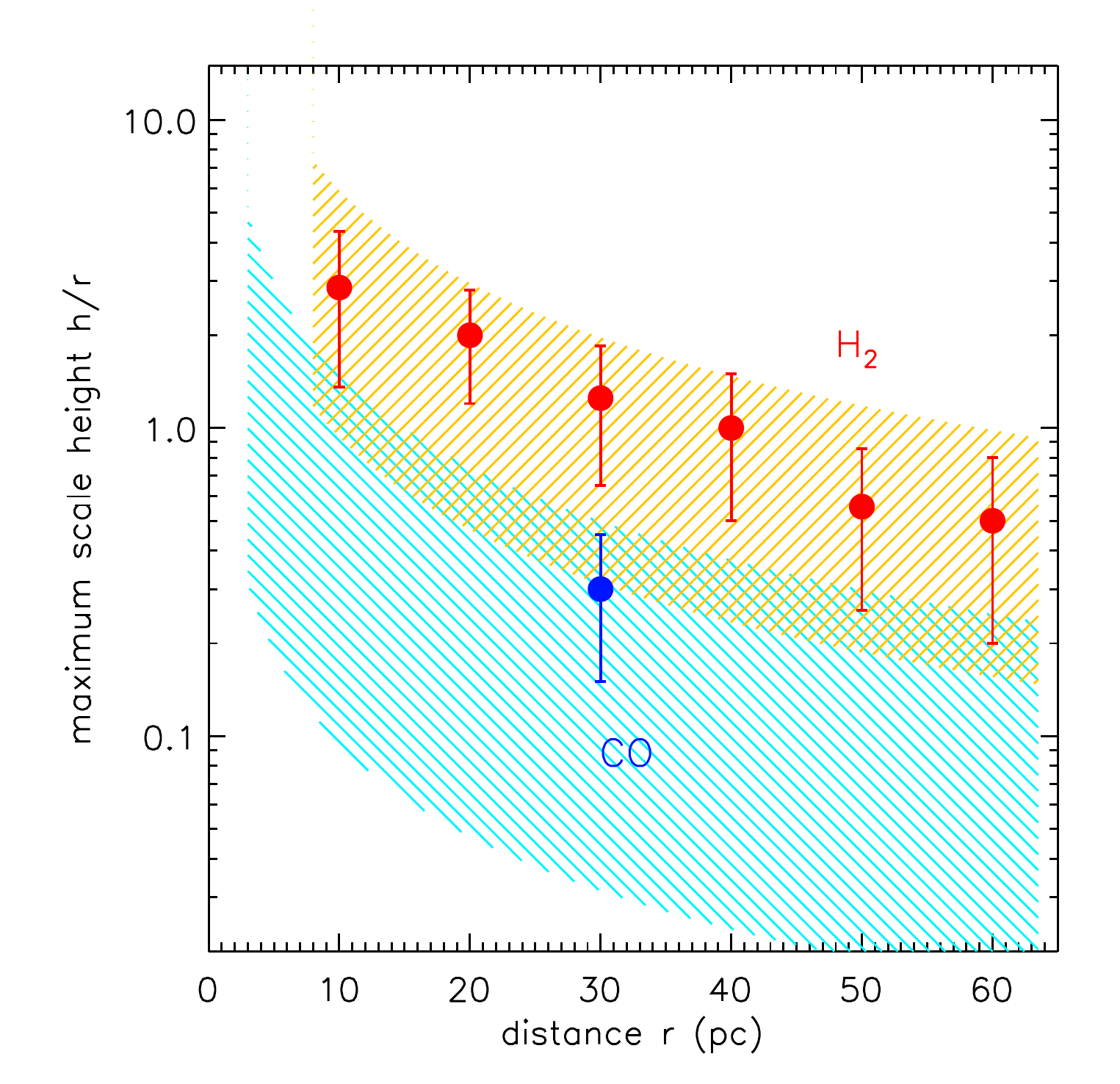}
\end{center}
\caption{\label{fig:max_hr}
Maximum achievable scale height for H$_2$-emitting gas clouds ($3.5 < \log n_H < 5$, $1000\,\mathrm{K} < T_\mathrm{gas} < 2000\,\mathrm{K}$; red-shaded area) and CO-emitting gas clouds ($5 < \log n_H < 7$, $20\,\mathrm{K} < T_\mathrm{gas} < 50\,\mathrm{K}$, blue-shaded area) for the collisional disk model. The red circles with errors show the approximate average of scale heights observed by \citet{Hic09} for H$_2$ while the blue circle with error bars indicates the observed CO scale height from Table~\ref{tab:scaleheight}. The inner cut-offs mark distances where the clouds at the respective densities cannot withstand the shear of the gravitational potential of the central black hole.}
\end{figure}

Fig.~\ref{fig:max_hr} shows the maximum scale height for clouds in the density ranges where the near-IR H$_2$ line observed by \citet{Hic09} and the sub-mm CO or HCN lines are emitted. The shaded theoretical areas are compared to averages inferred from \citet[][Fig.7]{Hic09} for H$_2$ and the range of CO scale heights listed in Table~\ref{tab:scaleheight}. For this, a volume filling factor of the gas of $\Phi_V=0.1$ was assumed with an H$_2$ gas mass fraction of 0.5 and a CO-to-H$_2$ mass ratio of 20\%. This shows that the observed density stratification leading to a multi-phase medium can be reproduced quantitatively by a collisional model. 

Note that even though the scale height of the H$_2$-emitting clouds can become geometrically thick, the maximum Hydrogen column densities per cloud implied from the Jeans radius and density criteria is $N_{H;\mathrm{max}} < 10^{22}$\,cm$^{-2}$. Given the drop of the radial density profile with radius (discussed in the following section), the number of clouds along a typical line of sight at distances $\ga$10\,pc at inclination 45$^o$ is $\ll$1. Together with the low column density, the geometrically thick part of the H$_2$ emitting gas is not expected to have a major role in obscuring the AGN, i.e. it can hardly count as \textit{``the torus''} (see also Sect.~\ref{subsec:uni_obsc}).

\subsubsection{The geometrically thin molecular disk: radial density profile of molecular lines}\label{subsubsec:thinmoldisk_rad}

The radial density structure of an accretion disk can be inferred from Poisson's equation,
$$\frac{\pdv g_r}{\pdv r} = -4\pi G \rho(r)$$
where $g_r = -\pdv \Phi(r)/\pdv r$ is the acceleration due to the gravitational potential $\Phi(r)$, which combines the various mass contributions, i.e. black hole, stellar cluster and the disk mass. For a radially non-self-gravitating disk, the density will follow $\rho(r) \propto r^{-1\ldots-3}$, with index $-3$ if the potential is dominated by the black hole and approximately $-1$ for domination by a stellar cluster \citep[e.g.][]{Bec04}.

The implication of this profile is that one can expect higher densities to be more prominent at smaller distances from the AGN. However, disks may be clumpy, as discussed in the previous section, in which case the density profile corresponds to the number density of clouds rather then the gas density of the clouds themselves, which in this case determine the emission from a particular region in the disk. Aside from the Jeans limit discussed above, clouds in a gravitational potential are also limited by shear and can be disrupted by tidal forces. Based on the Roche limit, clouds need to be denser than $n_H > 3 M_\mathrm{BH}/(8\pi m_\mathrm{gas} r^3)$. For a black hole with $M_\mathrm{BH}=10^7\,M_\sun$, clouds in the H$_2$-emitting density range will only be able to exist at distances $\ga8$\,pc from the AGN while CO-emitting clouds can reach down to $\sim3$\,pc (limits shown in Fig.~\ref{fig:max_hr}), unless these clouds are stabilised by some mechanism (e.g. magnetic fields) or the gravitational fields is shallower (e.g. due to dominance of a stellar cluster). Hence, observed disk sizes differ amongst different molecular emission lines and are expected to depend on typical density, with the highest density tracer lines showing the most compact emission. Indeed, the observed H$_2$O maser disks in Circinus and NGC1068 reach to sub-parsec scales, at which shear-stable densities are $n_H \ga 10^{10}$\,cm$^{-3}$, conducive to such maser emission.

Another consequence of molecular clouds becoming shear-instable at $\la$10\,pc is that the dusty molecular gas in this region is expected to become more smoothly distributed, with most substructure being filamentary or transitional in nature. This is consistent with radiation-(magneto-)hydrodynamic simulations \citep[e.g.][]{Dor12,Wad12,Dor17,Cha16,Cha17,Wil19}. In such an environment, the assumption of extreme clumpiness with very low volume filling factors and the treatment of the material in a purely stochastic framework breaks down. This affects commonly used radiative transfer models as well as the collisional disk model that provides the scale height via (elastic) cloud scattering. Therefore, at $r<10$\,pc, the disks are expected to flatten significantly as opposed to what is implied by the rising maximum $h/r$ in Fig.~\ref{fig:max_hr} at these distances.

\subsubsection{Geometrically thick obscuration and wind launching via IR radiation pressure}\label{subsubsec:windlaunch}

Thermal pressure is not able to sustain a thick disk over 10s of parsec scales, nor do high-density tracer lines show evidence for large column densities at large scale heights. Therefore, angle-dependent obscuration needs to occur on smaller scales. In addition, at $r<10$\,pc, collisional support is not viable as the medium becomes relatively smooth, so that other mechanisms are required to provide geometrical thickness (see previous section).

Towards the inner rim of the dusty disk, radiation pressure from the AGN gets stronger, with the radiation pressure force being $F_\mathrm{rad} \propto r^{-2}$. Radiation pressure on dust has been discussed as a potent mechanism to drive winds \citep[e.g.][]{Pie92,Hon07,Fab09,Ric17}, but most work focuses on the effects of the central AGN UV/optical radiation. \citet{Kro07} noted that, as the hot dust is optically thick to its own emission, near-IR radiation pressure can cause the inner rim of a dusty disk to puff up, providing some obscuration.

A dynamical analysis of the effect of IR radiation pressure is presented in \citet{Kro07}. In the present work, a simplified 1-dimensional description of the vertical structure in the innermost, hottest dust disk will be developed to demonstrate the effects of IR radiation pressure. The main interest lies in the IR radiation pressure's ability to increase the scale height of the otherwise thin disk. Therefore, near-IR radiation pressure is treated as a local pressure term. Following a similar description as for the pressure balance within stars, the near-IR radiation pressure can be written as $\mathrm{d}P_\mathrm{NIR}/\mathrm{d}r = \kappa_\mathrm{NIR}L_\mathrm{NIR}\rho(z)/(4\pi c r)$, with $L_\mathrm{NIR}$ being the near-IR luminosity of the dusty molecular disk and $\kappa_\mathrm{NIR}$ denoting the (Rosseland) mean opacity of the dust in the near-IR. The hydrostatic equilibrium (eq.~(\ref{eq:hse})) then leads to
$$ c_s^2\frac{\pdv\rho}{\pdv z} = -g_z \rho(z) \left(1-\ell_\mathrm{Edd} C_\mathrm{NIR} \frac{\kappa_d}{\kappa_g}\right) $$
where $\ell_\mathrm{Edd}$ is the Eddington ratio, $C_\mathrm{NIR}$ is the fraction of near-IR luminosity as compared to the AGN luminosity (= NIR ``covering factor''), $\kappa_d/\kappa_e$ is the ratio of near-IR opacity $\kappa_d$ and the gas opacity $\kappa_g$, and the sound speed $c_s = \sqrt{k_B T_\mathrm{gas}/m_\mathrm{gas}}$. This differential equation is very similar to the standard case of isothermal gas, with the scale height of the dusty molecular gas $(h/r)_d$ being a modified version of the scale height for isothermal gas $(h/r)$, 
\begin{equation}\label{eq:hseir}
    (h/r)_d = h/r \cdot \left(1-\ell_\mathrm{Edd} C_\mathrm{NIR} \frac{\kappa_d}{\kappa_g}\right)^{-1/2}.
\end{equation}
Gas in the sublimation region ($T_d\sim1500$\,K) has a temperature of up to $T_g\sim10^5$\,K (see Sect.~\ref{subsec:pheno_ir}). Accordingly, the scale height without dust would be $h/r \sim 0.04$ for a $10^7\,M_\sun$ black hole. On the other hand, a Seyfert AGN with $\ell_\mathrm{Edd}\sim0.05$ and a typical $3-5\,\micron$ bump with $C_\mathrm{NIR}\sim0.2$ \citep[e.g.][]{Mor12} will have $(h/r)_d \rightarrow \infty$ for dust grains with size $a<0.1\,\micron$, i.e. the IR radiation pressure will blow away such particles unless they are shielded. 

When interpreting the opacity as the area per unit mass of a dusty gas cloud, $\kappa_{dc} = 1/(N_\mathrm{H} m_\mathrm{gas})$, then $\kappa_{dc}/\kappa_g = 1/(N_\mathrm{H} \sigma_T)$, with $\sigma_T$ denoting the Thomson cross section. Dusty gas clouds with a column density $N_H\sim10^{23}$\,cm$^{-2}$ are settled in the disk with a similar scale height as for gas. However, gas clouds with several $N_H\sim10^{22}$\,cm$^{-2}$ will be dominated by the IR radiation pressure and puff up to high scale heights. As such column densities correspond to optical depths of $\tau_V\ga10$, these clouds will obscure the AGN and give the appearance of an optically thick absorber. The corresponding optical depth in the near-IR is $\tau_\mathrm{NIR}\sim1$, which are conditions favourable to dominating the near-IR emission. Accordingly, it is expected that the puff-up region of the disk dominates the observed $3-5\,\micron$ bump.

It is important to point out that once dusty gas becomes vertically unbound (i.e. $h/r \rightarrow \infty$), it will eventually be exposed to the radiation from the AGN. As the optical/UV luminosity of the AGN $L_\mathrm{UVO} > L_\mathrm{NIR}$ and $\tau_V \ga\tau_\mathrm{NIR}$, unshielded material lifted above the disk's mid-plane will experience strong radially outward radiation pressure, leading to the generation of a dusty wind, as shown by radiation-hydrodynamical simulations. More detailed dynamic analyses of this scenario are presented elsewhere \citep[e.g.][]{Kro07,Ven19}. Here, the important point is that while the near-IR radiation pressure will puff up the disk, it will not dominate the dynamics of the uplifted material once exposed to the AGN radiation, i.e. eq.~(\ref{eq:hseir}) illustrates the near-IR radiation's ability to make the disk geometrically thick, but any interpretation of the resulting structure beyond this should be cautioned.

To summarize, the near-IR radiation observed in the $3-5\,\micron$ bump is sufficient to puff up the dusty molecular disk in its innermost region and produce conditions favourable for launching a dusty wind driven by the optical/UV radiation from the central AGN. Black-body-equivalent temperatures of the $3-5\,\micron$ emission suggest that the diameter of this region is $\sim5-10\,\rsub$, corresponding to about $0.2-1$\,pc in nearby AGN.

\section{Discussion: Unifying molecular and IR continuum observations}\label{sec:unify}

The previous sections showed the seemingly different pictures emerging in the IR and sub-mm and presented a simple framework to assess the physical conditions that dominate at the (sub-)parsec scales and tens of parsec scales. Consequences of disk turbulence and radiation pressure point towards a multi-phase, multi-density structure forming around the AGN, as opposed to a literal interpretation of a ``dusty torus'' as a single physical entity. Indeed, the original cartoon depicting the obscurer in \citet{Ant85} does not allude to a large-scale geometric torus.

\begin{figure*}
\begin{center}
\includegraphics[width=1.78\columnwidth]{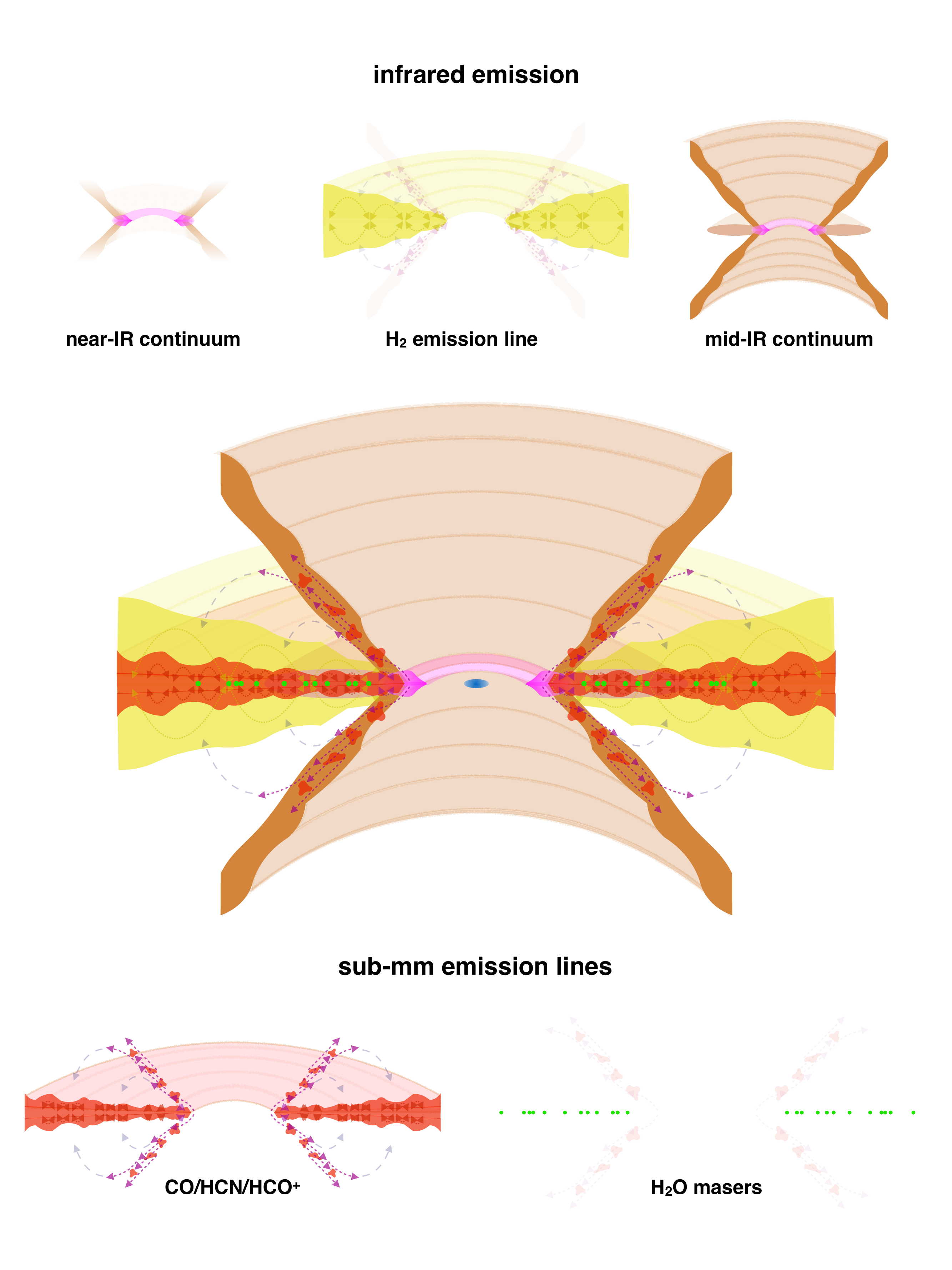}
\end{center}
\caption{\label{fig:multiphase}
Schematic view of the multi-phase dusty molecular environment of an AGN. The central picture has all empirically identified components plotted on top of each other (see Secs.~\ref{sec:ir_cont_view} \& \ref{sec:mol_view}). The top row shows those components of the multi-phase structure that can be detected by IR continuum observations (top left: near-IR, top right: mid-IR) and near-IR H$_2$ emission lines (top-middle panel). The bottom row shows the view in commonly observed sub-mm molecular lines (bottom-left) as well as by the H$_2$O maser emission. The arrows indicate the kinematics of the respective emission lines (without the rotational component).}
\end{figure*}

Fig.~\ref{fig:multiphase} shows a schematic of all the observed phases plotted on top of each other (see Sect.~\ref{sec:ir_cont_view} \& \ref{sec:mol_view}, with the colours matching those of Figs.~\ref{fig:uv_ir} \& \ref{fig:uv_submm}. In addition, arrows denote the dynamics of the respective component, combining the observed kinematics and theoretical expectations (see Sect.~\ref{subsec:phys_pic}). Two-sided arrows indicate turbulent motion, either due to cloud scattering, thermal processes, or IR radiation pressure, while single-sided arrows mark wind/outflow motion primarily due to AGN radiation pressure. The schematic contains similar elements as the one inferred by \citet{Ric17} based on the X-ray column-density distribution. This is not a surprise as the same fundamental physical effect -- radiation pressure on dusty gas -- has been invoked to explain those X-ray observations.

The following will sort the multiple phases into four regions that are distinct by the physical mechanism that dominate their shape and/or dynamics, based on the physical description in Sect.~\ref{sec:phys_desc}. Some consequences and relation to other phenomena are discussed.

\subsection{The cold, outer part of the equatorial disk}\label{subsec:uni_colddisk}

Summarising the results from Sects.~\ref{subsubsec:thinmoldisk_vert} \& \ref{subsubsec:thinmoldisk_rad}, at scales of $\ga5-10$\,pc, the dusty molecular mass is settled into a disk with a vertical density stratification. The dust temperatures in this region is relatively low (few 10\,K) as (self-)absorption of the central AGN radiation and its probable anisotropic radiation profile do not supply significant heating. Molecular excitation is dominated by collisions in a medium with high density contrast. 

The cold disk in a relatively shallow gravitational potential is prone to disturbances by host-galaxy conditions. For example, resonances, bars, bulge/pseudo-bulge dynamics, central star formation, or misalignment of the host galaxy with the plane of accretion can induce warps or cut-offs to the mass distribution. In such a potential, it is not necessarily expected that the rotational velocity component follows a Keplerian profile of a point-like mass, $v_\mathrm{rot} \propto r^{-1/2}$. This will mostly affect the larger scales of the cold-disk mass distribution ($\sim50-100$\,pc). However, even on the 10\,pc scale with a massive black hole dominating the gravitational potential, the rotational kinematics can appear non-Keplerian as the radiation pressure from the central AGN will flatten the effective potential. Given the vertical density profile, it is expected that lower density tracer lines will have lower $v_\mathrm{rot}$ than higher density tracer lines emitted from the same AGN distances as the corresponding gas is lifted higher above the mid-plane, thus being more exposed to the AGN radiation. 

On the larger scales, the cold disk can be considered the component linking the inner host galaxy to the accretion structure around the black hole. The total mass on tens of parsec scales is at least a significant fraction of the black hole mass \citep[e.g. $\sim5-10$\% in NGC3227, $\sim100$\% in NGC5643][]{Alo18,Alo19}, hence providing the mass reservoir for black hole growth. However, free-fall time scales for these distances are of the order of $10^7$ years, meaning that the observed gas masses are not directly linked to the current accretion/luminosity state of the AGN. Rather, they signify the mass available for the ongoing AGN duty cycle of the galaxy. The cold disk may therefore be considered the AGN feeding region. 

An important note to make is that while the feeding mass may originate from galactic scales, some form of dynamic processing needs to happen on scales of the (pseudo-)bulge or nuclear star cluster. If the gas would fall straight in, then one may expect that the rotation axis of the AGN is aligned with the angular momentum direction of the host galaxy. It is well established, however, that the direction of the AGN rotation axis is randomly distributed with regards to the host galaxy axis \citep[e.g.][]{Kin00}.

\subsection{The hot inner part of the equatorial disk}\label{subsec:uni_hotdisk}

In typical Seyfert-type AGN, scales of $<$10\,pc are wholly within or close to the sphere-of-influence of the supermassive black hole. In this region, the potential well will be steeper, leading to stronger sheer forces, a smoother medium, as well as a more centrally-concentrated mass distribution (see Sect.~\ref{subsubsec:thinmoldisk_rad}). Incidentally, at the same spatial scales, the energy density of the AGN radiation field will be stronger than the stars within the same volume \citep[e.g.][]{Jen17}. 

The inner ``hot'' part of the dusty molecular gas distribution is defined by sublimation of dust at $T_\mathrm{dust}\sim1\,500-2\,000$\,K. As shown in Sect.~\ref{subsubsec:windlaunch}, the dynamics of the dusty gas in this region will be strongly influenced by AGN and IR radiation pressure, leading to an inflation of the inner $1-5\,\rsub$ and ideal conditions to launch a wind. Such puff-up regions are not unique to AGN, but also seen in other accreting systems with significant amounts of dust around them, e.g. young stellar objects \citep[e.g.][]{Nat01,Dul01}. This puffed-up disk/wind-launching region will dominate the near-infrared emission of the AGN, creating the $3-5\,\micron$ bump seen in unobscured AGN. The observed covering factor of $\sim15-30$\% of the puff-ed up region implies that the hot disk contributes to the angle-dependent obscuration of the AGN, which will be further discussed in Sect.~\ref{subsec:uni_obsc}. 

As the puff-up/wind launching region of the disk is confined to the inner few $\rsub$, some AGN emission will penetrate through and illuminate the surface of the cooler parts of the disk. This is seen in IR interferometry as the sub-dominant mid-IR disk in some nearby AGN \citep[e.g.][]{Tri14}. In a disk with a vertical density profile, the emission is primarily emerging from its ``mid-IR surface'' (see principle 1 in Sect.~\ref{subsec:phys_cond}) and, given the steep radial profile, will naturally appear more compact than the wind region.

In Sect.~\ref{subsec:ir_intf}, it was pointed out that the hot dust region is dominated by emission from large graphite grains. As the dusty winds are launched in this region, one may expect that the parsec-scale dusty winds will be dominated by the same grain chemistry and size distribution. This will naturally lead to a reduction in observed silicate emission features as the mid-IR emission is dominated by the wind. Radiative transfer models taking into account differential dust destruction (i.e. silicate/graphite chemistry and grain sizes) and wind launching from the hot-dust region are able to reproduce the observed small near-IR sizes as well as the shallow silicate features \citep{Hon17,Gar17}. Further evidence for this scenario comes from extinction curves of type 1 quasars, which suggest a dearth of small grains in the outflow region \citep{Gas04}. At the same time, the wind region will be seen in both type 1 and type 2 sources, giving rise to the low anisotropy observed in the mid-IR (see Sect.~\ref{subsec:basics}). Notwithstanding the properties on parsec scales, it is well possible that on larger scales silicate dust may re-form in the wind \citep{Elv02} or that the ISM host galaxy may get entrained in the outflows. 

\subsection{The dusty molecular wind region}\label{subsec:uni_wind}

As described in Sect.~\ref{subsubsec:windlaunch}, IR radiation pressure will either increase the scale height of the hot disk (=``puff-up region'') or unbind dusty gas completely from the gravitational potential of the supermassive black hole. While this gas will experience a vertical pressure force, it will also be exposed to the AGN radiation. The AGN radiation pressure will be stronger than the near-IR radiation pressure by about a factor of $\tau_V/\tau_\mathrm{NIR}$, so that if dusty gas becomes unbound by IR radiation pressure and lifted upward, it will start to be radially pushed away from the AGN. Qualitatively, the shape is expected to be hyperbolic \citep[e.g.][]{Ven19}, as has been implied by radiative transfer modelling of IR interferometry \citep{Sta19}. This naturally leads to a hollow-cone configuration with dust confined towards the edges of the cone. A Similar configuration is also obtained assuming X-ray heating instead of radiation pressure \citep[e.g.][]{Bal93}, and is consistent with edge-brightened narrow-line regions seen in some AGN \citep[e.g.][]{Can03}.

In the Seyfert regime with $\ell_\mathrm{Edd} \sim 0.01-0.2$, wind launching and driving will be sustained by dusty gas with Hydrogen column densities of  $10^{22}-10^{23}$\,cm$^{-2}$. This corresponds to optical depths of $\tau_V \sim 10-100$. Therefore, the dusty outflow will contribute to obscuring the AGN (see Sect.~\ref{subsec:uni_obsc} for a more comprehensive discussion). As dusty gas being exposed to the AGN radiation may heat up and expand adiabatically, the highest columns are probably expected closer to the AGN. However, as radiation-hydrodynamical simulations show, dense optically-thick gas is lifted from sub-parsec scales to parsec scales and beyond, providing the required $\tau_\lambda\sim1$-surface and covering factors in the mid-IR to contribute or even dominate the mid-IR emission of AGN within the central 100\,pc \citep[e.g.][]{Wad16,Wil19}.

While most of the discussion focused on radiation pressure on dusty gas, magnetic driving in dusty gas has also been considered as a possible mechanism \citep[e.g.][]{Kon94,Cha17,Vol18}. However, these models often invoke radiation pressure to initiate or maintain the outflow, demonstrating that mechanisms causing dusty winds require at least some degree of radiation support.

\subsection{Mass outflow rate from dusty molecular winds}\label{subsec:mdotw}

Winds are a major factor in AGN feedback on galaxies and several driving mechanisms, including radiation driving, have been proposed. Hence, it is worth putting the parsec-scale dusty molecular winds into the context of feedback and determine typical mass-loss rates from these winds. Radiation pressure is a form of momentum-driving, so that the mass outflow rate of an optically-thick, spherical wind $\dot{M}_w \approx L_\mathrm{AGN}/ (v_\infty c)$, where $v_w$ is the terminal speed of the wind and $v_w\sim v_\mathrm{esc}=\sqrt{2GM/R}$ is approximately of the same order as the escape velocity at the radius $R=R_{\tau=1}$ where the wind transitions from optically-thick to optically-thin \citep{Mur05,Tho15}. Replacing the black hole mass $M=\kappa_g L /(4\pi \ell_\mathrm{Edd} G c)$ leads to
\begin{equation}\label{eq:mdotw_full}
    \dot{M}_w = \left(\frac{2\pi}{c \kappa_g}\right)^{1/2} \ \ell_\mathrm{Edd}^{1/2} \ L_\mathrm{AGN}^{1/2} \ R_{\tau=1}^{1/2}
\end{equation}
Evaluating eq.~(\ref{eq:mdotw_full}) requires an estimate for the radius where the wind becomes optically thin. A lower limit can be inferred from the significant contribution of the dusty winds to the mid-IR interferometry flux, which implies that the wind (or clumps/filaments in the wind) do have at least $\tau_\mathrm{12\,\micron}\sim1$, i.e. $\tau_V\sim10$, on parsec scales. Observationally, the observed 12\,$\micron$ emission sizes have been found to be of the order of few \,pc, mostly independent of AGN properties \citep[e.g.][]{Kis11b,Bur13}. Given the relatively small AGN parameter range probed by past IR interferometric samples, it is possibly that the 12\,$\micron$ sizes do nevertheless scale with $L$ or $\ledd$ \citep{Lef19}. An upper limit for $R_{\tau=1}$ can be obtain from the observation that some AGN do show low surface brightness polar features in single-telescope mid-IR images \citep[e.g.][]{Asm16,Asm19}. Those faint features on 10s to 100s pc scales may be emitted by optically-thin dusty gas, setting $R_{\tau=1}<10-100$\,pc. From radiative transfer modelling of mid-IR images of the Ciricnus galaxy, \citet{Sta17} derive $\tau_V\sim1$ out to 40\,pc. Based on these constraints, a normalisation of $R_{\tau=1}=5$\,pc may be considered a reasonable estimate in the absence of resolved mid-IR observations, but it may require adjustment if such constraints are available.

There are further factors of order unity that are not accounted for yet: First, eq.~(\ref{eq:mdotw_full}) assumes a spherical shell, while the dusty wind launching region has a covering factor of $C_\mathrm{NIR}\sim0.2$ (see Sect.~\ref{subsubsec:windlaunch}). Second, near-IR radiation from the disk and the optically-thick wind itself will boost the momentum transfer to the dusty gas by up to a factor of $b_\mathrm{NIR}\la10$ \citep[e.g.][]{Tho15,Ish15}. The resulting correction factor to eq.~(\ref{eq:mdotw_full}) is $f_c = b_\mathrm{NIR} C_\mathrm{NIR} \sim 0.8$ for $b_\mathrm{NIR}=4$ and $C_\mathrm{NIR}=0.2$. Accounting for this correction and setting $R_{\tau=1}=5$\,pc, eq.~(\ref{eq:mdotw_full}) can be re-written as
\begin{equation}\label{eq:mdotwind}
    \dot{M}_w = 2.5\,M_\sun/\mathrm{yr} \times L_{\mathrm{AGN};44}^{1/2}\ \ell_{0.05}^{1/2} \ R_{\tau=1;5}^{1/2}
\end{equation}
with $L_\mathrm{AGN}$ in units of $10^{44}$\,erg/s, $\ell_\mathrm{Edd}=0.05$, and $R_{\tau=1}=5$\,pc. 

The anticipated mass outflow rates are broadly consistent with order of magnitudes of observed small-scale nuclear outflows. As an example, \citet{Alo19} determine the mass outflow rate in NGC~3227 from ALMA CO observations within the inner 15\,pc around the AGN. They find $\dot{M}_w\sim0.6-5\,M_\sun$/yr for a typical Seyfert AGN with a luminosity of $L_\mathrm{AGN} \sim 2\times10^{43}$\,erg/s. The expected outflow rate from eq.~(\ref{eq:mdotwind}) would be $\dot{M}_w \sim 0.7\,M_\sun$/yr. Similarly, \citet{Zsc16} report a potentially AGN-driven molecular outflow in the Circinus Galaxy in the range of 0.35$-$12.3\,M$_\sun$/yr. Using the probable luminosity range and Eddington ratio of Circinus \citep[$\log L($erg/s$)=43.36-44.43$, $\ledd=0.2$][]{Sta19} implies that AGN radiation pressure on dusty gas can drive an outflow of 1.7$-$10.5\,M$_\sun$/yr, consistent with the observations. On the higher-luminosity end, \citet{Aal15} presented ALMA HCN and HCO$^+$ observations of Mrk 231. For their potentially highly accreting AGN ($\ell_\mathrm{Edd}\sim0.5$) with a luminosity $L_\mathrm{AGN}\sim10^{46}$\,erg/s, an outflow rate of $\dot{M}_w\sim250\,M_\sun$/yr is estimated from eq.~(\ref{eq:mdotwind}). The authors report a mass outflow rate as $80-800\,M_\sun$/yr, consistent with the presented model of a dusty wind being driven off the pc-scale environment. 

Comparing the outflow rate on ``torus'' scales to the accretion rate on accretion-disk scales, $\dot{M}_\mathrm{acc} = L_\mathrm{acc}/(\eta c^2)\sim0.02\,\mathbf{M_\sun/\mathrm{yr}}\cdot L_{\mathrm{AGN};44}$, with $\eta\sim0.1$ being the accretion efficiency, shows that $\dot{M}_w$ exceeds the mass supply to the black hole through the inner accretion disk by a factor of 100. This is not too surprising as observed mass inflow rates of cold molecular gas in Seyfert galaxies on tens to hundreds of parsecs scales are of the order of $1-10\,M_\sun$ \citep[e.g.][]{Sto19}, exceeding the central accretion rate by a similar factor. As such, mass conservation requires the presence of massive outflows on parsec scales \citep[see also][]{Eli06}, probably combined with a fraction of the inflowing mass lost to star formation\footnote{On scales of $<$50$-$100\,pc, it is difficult to estimate the star formation rate as many tracer lines experience contamination from the AGN \citep[e.g.][]{Jen17}.}.

Please note that the meaning of outflow in the context of this paper refers to the parsec to hundreds of parsec scales, i.e. escape from the sphere of influence of the black hole. It is well possible that some of these outflows will not escape the host galaxy but will deposit or distribute the entrained mass within the bulge or galaxy. The outflow rates are significant enough to consider dust-driven molecular winds from near the torus region as a mode of AGN feedback and mechanism for self-regulation of black hole growth.

While the present work focuses on radio-quiet Seyfert-type AGN, molecular outflows are also observed for radio-loud \citep[e.g. 3C273][]{Hus19a} or radio-intermediate AGN \citep[e.g. IC5063, HE 1353-1917][]{Das16,Oos17,Hus19b}. In these objects, the molecular outflows are often collimated and co-spatial with the jet. While energetical and momentum arguments do not always allow for unequivocally pinning down the dominating physical mechanism \citep[e.g.][]{Das16,Hus19b}, it is likely that those outflows are driven by a jet mode rather than radiation pressure. However, the presence of a jet by itself is not always a clear discriminator between jet- or radiation-driven outflows \citep{Wyl18}.

\subsection{The dark fall-back region}\label{subsec:uni_fallback}

As the dusty molecular wind is optically thick to the AGN radiation at least close to the wind launching region, it will provide some obscuration (see Sect.~\ref{subsec:uni_obsc}). In order to sustain wind driving over some time, dusty gas must remain exposed to the AGN radiation near the launching region for maximum momentum deposition (see Sect.~\ref{subsec:mdotw}). Gas being swept up in a region that becomes self-obscured to the AGN radiation will have less momentum deposited, giving rise to a ``failed wind''. Such mass fallback has been seen in radiation-hydrodynamic simulations \citep[e.g][]{Wad12,Wil19}, with observational evidence also found in low-velocity molecular outflow components of the Ciricnus galaxy \citep{Izu18}. Gas presence in this fall-back region is transitional, probably of lower column density, and the lack of direct AGN radiation makes the region dark in the IR. The molecular gas in this region is co-spatial with the hotter, lower-density gas of the equatorial disk with which it may interact dynamically. As such, it is difficult to pin down the exact physical properties of gas falling back. 

The simulations by \cite{Wad12} suggest that the fall-back material may induce strong turbulence in the disk, making it geometrically thick. However, other models do not find this effect \citep[e.g.][]{Dor12,Cha16,Wil19} as the shocked gas in the disk rapidly cools radiatively.

\subsection{Reproducing toroidal obscuration}\label{subsec:uni_obsc}

The original reason to postulate the torus was the observed angle-dependent obscuration of AGN. However, toroidal obscuration may be caused by a range of mass distributions with circo-symmetric geometry and does not imply geometrical thickness over a large radial range. Indeed, \citet{Ant85} only postulate geometrical thickness as a requirement, without any preference as to how the optically-thick mass is distributed. The structure discussed in this present paper does provide the required angle-dependent obscuration. 

In the structure proposed here, the highest column densities, probably exceeding the Compton-thick limit with optical depth $\tau_V\ga1000$, are encountered when viewing the equatorial disk edge-on. As discussed in Sec.~\ref{subsubsec:thinmoldisk_vert}, the scale height of this region for the dense gas and $N_H\ga10^{23}$\,cm$^{-2}$ as seen in CO is typically below $\sim0.15-0.3$, with lower density gas reaching higher. The wind and wind-launching regions provide additional obscuration with higher covering factors. While the puff-up in the inner hot disk will have covering factors of $C_\mathrm{NIR}\sim0.2-0.3$, Sect.~\ref{subsubsec:windlaunch} discusses that, depending on $\ell_\mathrm{Edd}$, dense clouds will be elevated and driven away by the dusty wind. For the range of typical Seyfert Eddington ratios of $\ell_\mathrm{Edd}=0.01-0.2$, up-lifted dust clouds/filaments will have column densities of the order of $N_H=10^{22}-10^{23}$\,cm$^{-2}$ (see Sect.~\ref{subsec:uni_wind}). Hence, the wind will contribute to the obscuration of the AGN. The opening angle of the wind, therefore, delineates the obscured from the unobscured region and sets the observed covering factor.

With the wind contributing to obscuration, it can be expected that a subset of AGN will be viewed close to the edge of the hollow cone, with dense gas moving towards the observer. This outflowing material may be related to the warm absorber seen in some type 1 AGN in the X-rays, \citep[e.g.][]{Tur93,Ric10}, with NGC~3783 being such a candidate object with a warm absorber where modelling of the mid-IR interferometry implies a viewing angle along the edges of the outflow cone \citep{Hon17}.

\subsection{Relation to the accretion structure inside the dust sublimation radius}\label{subsec:inside_rsub}

The presented picture of the mass distribution around the AGN is limited to the structure from tens to hundreds of parsec scales down to the sublimation region, with the physics being linked to specific properties of dusty molecular gas. It is interesting that a similar structure emerges when considering the primarily dust-free\footnote{Note that it has been suggested that a small amount of dust may still be present inside the observed sublimation radius that could control some of the gas dynamics \citep[e.g.][]{Cze11,Bas18}.}, atomic and ionised gas phase inside the sublimation radius. \citet{Elv00} discuss the distribution of gas leading to broad and narrow absorption/emission lines in quasars. Similarly to the present work, radiation forces on the gas via its continuum and line opacity cause a wind to emerge from the accretion disk. Most of the absorption line phenomena are attributed to different phases of this wind. The dusty winds discussed in the present manuscript probably define the boundary to the outflows emerging from the dust-free region. While their velocities can be much higher, considering mass conservation, the mass load of the dusty winds discussed in Sect.~\ref{subsec:mdotw}, and the resulting reduced accretion rate in the accretion disk region as compared to the dusty region, these flows supposedly escape close to the skin of the dusty winds.

\section{Conclusions}\label{sec:conc}

This paper aimed at reviewing the general properties of radio-quiet Seyfert-type AGN as seen in the infrared (IR) and sub-mm on scales $<$100\,pc. Those scales refer to the dusty molecular environment commonly referred to as the ``torus''. The observations in both wavelength regimes have been unified with a simple set of physical principles, drawing the picture of a multi-phase, multi-component region. The major conclusions are as follows:
\begin{itemize}
    \item The dusty molecular gas flows in from galactic scales of $\sim$100\,pc to the sub-parsec environment (with the sublimation radius $\rsub$ as the inner boundary) via a disk with small to moderate scale height. Higher density gas in radial direction is observed closer to the AGN and in vertical direction closer towards the mid-plane.
    \item The disk puffs up within $\sim$5\,$\rsub$ of its inner edge due to IR radiation pressure. In this region, gas with column densities $N_H\la10^{22}-10^{23}$\,cm$^{-2}$ becomes unbound and is swept out in a dusty wind by radiation pressure from the AGN.
    \item The radiation-pressure-driven dusty molecular wind carries significant amounts of mass. The $\sim$pc-scale wind outflow rate is estimated as $$\dot{M}_w = 2.5\,M_\sun/\mathrm{yr} \times L_{\mathrm{AGN};44}^{1/2}\ \ell_{0.05}^{1/2} \ R_{\tau=1;5}^{1/2},$$
    which is broadly consistent with molecular outflows seen by ALMA on these scales. Such rates can explain the difference of a factor $\sim$10 between galactic-scale inflow rates onto AGNs and the small-scale accretion rates from the accretion disk. Interestingly, for a given black hole mass, $\dot{M}_w \propto L \cdot R_{\tau=1}$. If the sizes of the dusty winds do increase with luminosity (or Eddington ratio), then higher luminosity AGN will remove a lager fraction of the inflowing gas than their lower luminosity counterparts, thus limiting their own mass supply towards the black hole. Therefore, dusty molecular winds are a mechanism to self-regulate AGN activity and will provide feedback from the AGN to the host galaxy.
    \item Angle-dependent obscuration is caused primarily by the cool disk (circumnuclear $N_H\ga10^{24}$\,cm$^{-2}$) as well as the wind-launching region and hollow-cone wind $N_H\sim10^{22}-10^{23}$\,cm$^{-2}$. Hence, even when defining the ``torus'' simply as the obscurer of the AGN, it will still consist of multiple spatial and dynamical components rather than a single entity.
\end{itemize}
It is important to point out that the picture drawn in this paper is derived from the similarities shared by the various AGN observed with IR interferometry and in the sub-mm. Individual sources will show some degree of deviation from this picture, specifically on the 10s parsec scales, as orientation of and interaction with the host galaxy and varying degrees of nuclear starformation will affect the mass flow. In the framework of radio-quiet, local Seyfert-type AGN, it is consistent with the proposed structure to explain X-ray obscuration of similar type of AGN \citep{Ric17}.

\acknowledgements
\textit{Acknowledgements} --- The author wants to thank C. Ramos Almeida who inspired this paper by her inviting the author to give a talk on this topic at the EWASS 2019 Session ``The ALMA view of nearby AGN: lessons learnt and future prospects.'' Further, the author is thankful to Ski Antonucci for many insightful comments and suggestions, D. Williamson for discussions on the outflow properties, and P. Gandhi for input from the X-rays. This work was supported by the EU Horizon 2020 framework programme via the ERC Starting Grant \textit{DUST-IN-THE-WIND} (ERC-2015-StG-677117).

\end{document}